\documentclass[useAMS,usenatbib]{mn2e}

\usepackage{graphicx}
\usepackage{subfigure}
\usepackage{caption}
\usepackage{epsfig}
\usepackage{appendix}
\usepackage{amsmath}
\usepackage{arydshln}
\usepackage{multirow}
\usepackage{natbib}

\title[Bursting emission from B0611$+$22]
{Bursting emission from PSR B0611$+$22}
\author[A. D. Seymour et al.]{A.~D.~Seymour$^{1}$\thanks{E-mail:
aseymour@mix.wvu.edu},  D.~R.~Lorimer$^{1,2,3}$ and
J.~P.~Ridley$^{4}$\\
$^1$Department of Physics, West Virginia University, Morgantown, WV 26505, USA\\
$^2$National Radio Astronomy Observatory, Green Bank, WV 24944, USA\\
$^3$Astrophysics, University of Oxford, Denys Wilkinson Building, Keble Road, Oxford OX1 3RH\\
$^4$Department of Engineering and Physics, Murray State University, Murray, KY 42071, USA
}

\begin{document}
\date{Accepted; \today}
\pagerange{\pageref{firstpage}--\pageref{lastpage}} \pubyear{2013}

\maketitle

\label{firstpage}

\begin{abstract}
Over the past decade it has become apparent that a class 
of `bursting pulsars'  exist with the discovery of PSR~J1752+2359 and PSR~J1938+2213. 
In these pulsars, a sharp increase in the emission is observed
that then tends to systematically drop-off from pulse-to-pulse. In this paper we describe the 
discovery of such a relationship in high-sensitivity observations of the young (characteristic age of $90,000$~yrs) 0.33~s pulsar B0611+22
at both 327~MHz and 1400~MHz with the
Arecibo radio telescope. While Nowakowski previously showed that B0611+22 has mode-switching properties, 
 the data presented here show, for the first time, that this pulsar emits
 bursts with characteristic time-scales of several hundred seconds.
 At 327~MHz, 
the pulsar shows steady behaviour in one emission mode which is
enhanced by bursting emission slightly offset in pulse phase from this
steady emission. Contrastingly at 1400~MHz, the two modes
appear to behave in a competing operation while still offset in phase. Using a fluctuation spectrum analysis,
 we also investigate each mode independently
for sub-pulse drifting.  Neither emission mode (i.e.~during bursts or
persistent emission) shows the presence of the drifting sub-pulse phenomenon.
The bursting phenomena seen here appears to be a hybrid
between bursting seen in other pulsars and 
the bistable profile illumination behaviour reported
in two other pulsars by Rankin et al. 
Further examples of this cross-frequency behaviour are required, as this phenomenon 
may be quite common among the pulsar population.

\end{abstract}

\begin{keywords}
methods: data analysis $-$ pulsars: general $-$ pulsars: individual: B0611$+$22
\end{keywords}

\section{Introduction}

High-precision timing observations which probe neutron star physics,
general relativity, and other phenomena rely on tracking rotational
phase as measured by template matching of stable integrated pulse profiles
\citep[for a discussion, see][]{Lorimer:2005fk}. Upon closer examination,
however, the radio emission displays a rich variety of examples in
which pulsars appear to flip between various states on a wide range
of timescales. 

Shortly after the discovery
of pulsars, in a remarkable series of papers making use of the 
Arecibo telescope, \cite{1970Natur.227..692B, 1970Natur.228...42B,1970Natur.228..752B,1970Natur.228.1297B}
 showed that pulsars exhibit
nulling, mode-changing and sub-pulse drifting phenomena in their
pulse-to-pulse emission. 
Nulling, the abrupt halting of the emission
mechanism for a few or many hundreds of pulses \citep{1970Natur.228...42B} can,
be viewed as an extreme version of mode-changing \citep{1970Natur.227..692B,1970Natur.228..752B,1970Natur.228.1297B} 
during which 
the pulse profile exhibits two or more stable pulse shapes.
 Although not universal, an analysis of pulse nulling across the population of isolated pulsars by \cite{2007MNRAS.377.1383W} tends to suggest
that the fraction of time spent in the null state increases with characteristic age.
Sub-pulse drifting \citep{1968Natur.220..231D,1973NPhS..243...77B},
where clearly ordered subpulses drift
across the main pulse window at a fixed rate characterized by 
a subpulse spacing and a repetition period appeared to be
preserved over an interval of time spent in the null state 
\citep{1978MNRAS.182..711U}.

With the passage of time, more discoveries have blended and questioned these classifications.
\cite{1970ApJ...162..727H} discovered the first of a number of pulsars that appear to combine 
mode changing with sub-pulse drift, where 
these pulsars generally assume one of several sub-pulse drifting modes.  

Also, \cite{MNR:MNR10512} have shown that some pulsars
exhibit what they term ``bistable profile illumination'', which is single-pulse drifting
that mimics the abrupt beginnings of mode-changing but then
gradually drifts into phase with the normal mode.  
 \cite{2007MNRAS.377.1383W} suggest
that nulling and mode changing are different manifestations of the same phenomenon.

Other extraordinary pulsars that may relate nulling to mode-changing occur on a wide 
range of timescales. A few of these discoveries have been
the rotating radio transients (RRATs; \cite{2006Natur.439..817M}), the intermittent pulsars 
\citep{Kramer28042006, 0004-637X-746-1-63, 2012ApJ...758..141L} and
the observation of bursting emission, which was
until now only seen in J1752$+$2359  \citep{0004-637X-600-2-905} and J1938$+$2213 
\citep{2003PhDT.........2C, Lorimer2013}. The bursting pulsars bear some resemblance
to conventional nulling pulsars, with the exception that
the pulse to pulse intensities are modulated by a decaying envelope before 
abruptly returning to a null or faint state.
While the RRATs emit only occasional
individual pulses spaced anywhere between seconds and years,
intermittent pulsars switch on and remain steady before switching off
again over timescales of days to years. Over these longer timescales,
where it is possible to measure the spin period derivative precisely,
it is observed that the ``on'' and ``off'' states for the intermittent
pulsars are associated with two different spin-down rates\citep{Kramer28042006}. 

A similar behaviour was recently seen in 17 non-intermittent
pulsars \citep{DataHome} which switch between different spin-down
states in a quasi-periodic or even chaotic fashion \citep{2013MNRAS.428..983S}, 
and are sometimes accompanied by correlated pulse profile changes.  
\cite{DataHome} suggest that the mixture of spin-down states could even account for the ``timing noise''
variations seen in many pulsars and suggest that most, if not all,
normal pulsars have multiple spin-down states indicating global
changes in magnetospheric current densities. One implication of these
results is that mode changing (and perhaps, by association, bursting)
could be connected to spin-down rate changes on much shorter timescales. 
Very recently, synchronous X-ray and radio state switching was reported in the 
classical mode-changing pulsar B0943+10~\citep{2013Sci...339..436H}.
In this case, non-thermal
unpulsed X-rays are observed during the bright radio emitting
phases, while an additional pulse thermal component is present
during the radio-quiet phase. 
A better understanding of the observational phenomenology involved is
now required in order to make further progress.

In this paper, we present a previously unseen bursting emission phenomenon in 
PSR~B0611$+$22 which was discovered by \cite{DAVIES:1972fk}. 
This young pulsar with a characteristic age of $90,000$~years, 
has considerably high timing noise \citep{1980ApJ...237..206H}.
This high level of noise could indicate that there is still some underlying 
phenomena that is not being accounted for in the timing models.
At first glance, it appears to be a normal pulsar with a simple pulse shape - essentially Gaussian in appearance
that is approximately 80\% linearly polarized \citep{1989ApJ...346..869R}. 
Yet, it has been shown that the integrated pulse shape varies at 
different times \citep{1980ApJ...239..310F}. 
 \cite{1992msem.coll..280N} has also shown that there is a correlation with the 
 brighter pulses and their phase alignment, and later claimed that the timing noise 
 was due to mode switching \citep{2000AAS...19713004N}.
 Other efforts have been made to find a relationship contributing to this timing noise, such as  
 sub-pulse drifts \citep{2007A&A...469..607W}, but to no avail.
 
 In this paper, using archival Arecibo observations, we have been able to 
 reveal that this pulsar exhibits bursting episodes and confirm that it has 
 moding behaviours.
 How this was performed and the implications are outlined in this paper in the following way. 
In Section 2 we provide technical information about 
the observations. In Section 3 we detail the data analysis techniques used to expose 
relationships from pulse-to-pulse. In Section 4 we conduct a fluctuation
spectrum analysis on the normal and bursting modes found in Section 3.
In Section 5, motivated by the recent X-ray observations of PSR~B0943+10,
 we make predictions for observations of B0611+22 at
X-ray wavelengths that may help to understand the phenomenon seen here.
Finally, in Section 6, we discuss our results and their scientific significance.

\section{Observations}

The data presented here were collected between March 2 and 8,
2009 (MJD range 54892--54898)
in a dedicated observing run with the Arecibo telescope to search for unusual
spectral index behaviour in PSR~B0611$+$22. To characterize the flux density
spectrum, observations were carried out at 327~MHz, 1400~MHz, 4.5~GHz, and 8.8~GHz
using the Wide Band Arecibo Pulsar Processors \citep[WAPPs; ][]{2000ASPC..202..275D}. 
Unfortunately, the pulsar was not detected at 8.8 GHz and radio frequency interference (RFI) 
dominated the 4.5 GHz observations. 
Further details on the individual observations in which the pulsar 
was clearly detected are shown in Table \ref{tab:Info}.

\begin{table*}

\begin{center}
\caption{Summary of Arecibo observations of B0611$+$22.}

\label{tab:Info} 

\begin{tabular}{ccccccc}
	\hline
	MJD & Centre Frequency  & Integration Time& Bandwidth/Channel & Sample time  & Number of Channels \\
	        & (MHz) & (s) &  (MHz) &  ($\mu$s) &  \\
	\hline
	
	54892.97 & 1400 & 900 & 0.195    & 256 &  2048\\
	54893.01 &  327  &  900 & 0.012    &  256 & 4096\\
	
	54894.95 & 327   &   900  & 0.012 & 256 & 4096\\
	54894.97 & 1400 &   900    & 0.781 & 128 & 512\\
	
	54896.94	 & 327   & 900 & 0.012&256 & 4096\\
	54896.97	&  1400& 900 & 0.781 &128 &512\\
	
	54898.95	&  1400& 900 & 0.781 &128 &512\\
	54898.97 & 327 & 900 &0.012& 256  &4096\\
	
	\hline

\end{tabular}
\end{center}

\end{table*}%

\section{Data analysis}

The data from each of the observations listed in in Table \ref{tab:Info}
were analyzed in a systematic way, as described below.

\subsection{Individual pulse profiles}

The autocorrelation functions recorded by the WAPPs 
in each polarization channel were summed and Fourier transformed 
to convert them to the equivalent set of total-power 
spectral channels using standard data analysis
techniques \citep[see, e.g., Section 5.2.2 of ][]{Lorimer:2005fk}
implemented in the {\tt filterbank} program as
part of the \textsc{sigproc} \footnote{http://sigproc.sourceforge.net/} 
pulsar processing package. The 
resulting data were then dedispersed at the pulsar's dispersion measure (96.91~cm$^{-3}$~pc) and folded
at the predicted pulse period (0.33 s) using the ephemeris provided in
the ATNF pulsar catalog \citep{2005AJ....129.1993M}\footnote{www.atnf.csiro.au/people/pulsar/psrcat/}.
The \textsc{sigproc} {\tt fold} program was used to produce
an ASCII pulse profile with 256 rotational phase bins. 
Then using the standard deviation of the off-pulse bins,
we were able to convert to mJy using the radiometer equation 
\citep[see, e.g., Section 7.3.2 of][]{Lorimer:2005fk}, with a gain of 10~K/Jy and an 
assumed system temperature of 122~K for 327~MHz and 30~K for 1.4~GHz.  
Zoomed-in versions of these profiles are shown in the left panels of 
Figs.~\ref{fig:98TBOX} and \ref{fig:98LBOX}.

\subsection{Revealing the emission modes}
	
When looking for a relationship across pulse intensities, there is
often a great deal of noise present in the system which may obscure
any underlying correlation. A simple way to reveal these relationships is to take a
boxcar convolution (a running average) over several data points. This
method increases the signal-to-noise ratio by the square root of the
total number of data points that were averaged, but one must be
aware that this will also reduce the resolution of the data along the
direction of the averaging. We carried out a boxcar convolution of
the data with a normalised area kernel along 64 data points in a
single bin, to preserve the bin resolution. We chose this kernel 
size to reduce the noise fluctuations
by at least a factor of five, and to have a multiple of two for computing
efficiency. This averaging is often known as smoothing, and we will
refer to these post-processed data as smoothed data. 
		
\begin{figure*}
\includegraphics[width=.75\linewidth]{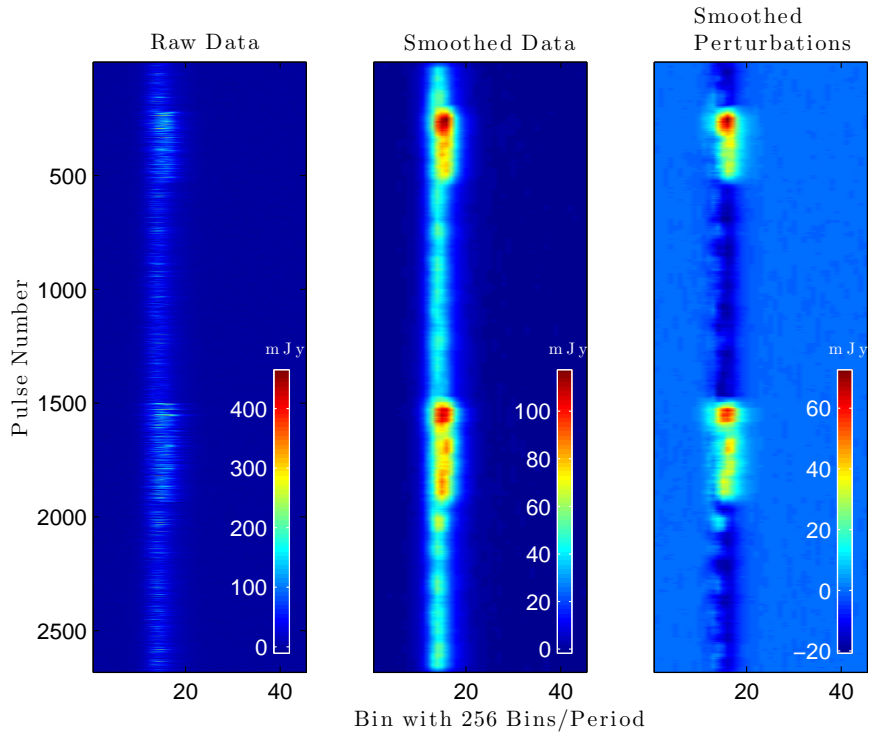}
\caption{The 327~MHz observation of PSR~B0611$+$22 on MJD 54898. 
         Left: the folded dedispersed time series. Centre: 
	 the time series after being box-car convolved along each bin 
         with a kernel of 64 pulses. Right: the convolution of the 
         perturbations about the mean profile with the same kernel.} 
	\label{fig:98TBOX}
\end{figure*}
								
When smoothing is performed on the 327~MHz emission 
from PSR B0611$+$22, the mode changes suggested in \cite{1992msem.coll..280N} 
are readily apparent, which is shown in the centre panel of Fig.~\ref{fig:98TBOX}. 
There, two prolonged episodes of
enhanced emission can be seen. These enhanced emission 
episodes can then be isolated when the average 
of each bin is subtracted from that bin to form the perturbations 
about the mean. These perturbations are then smoothed to produce the 
right panel of Fig.~\ref{fig:98TBOX}. It appears that 
the enhanced episodes are additional effects overlying a normal emission mode.

\begin{figure}
{\includegraphics[width=\linewidth,trim=25 0 100 0,clip=true ]{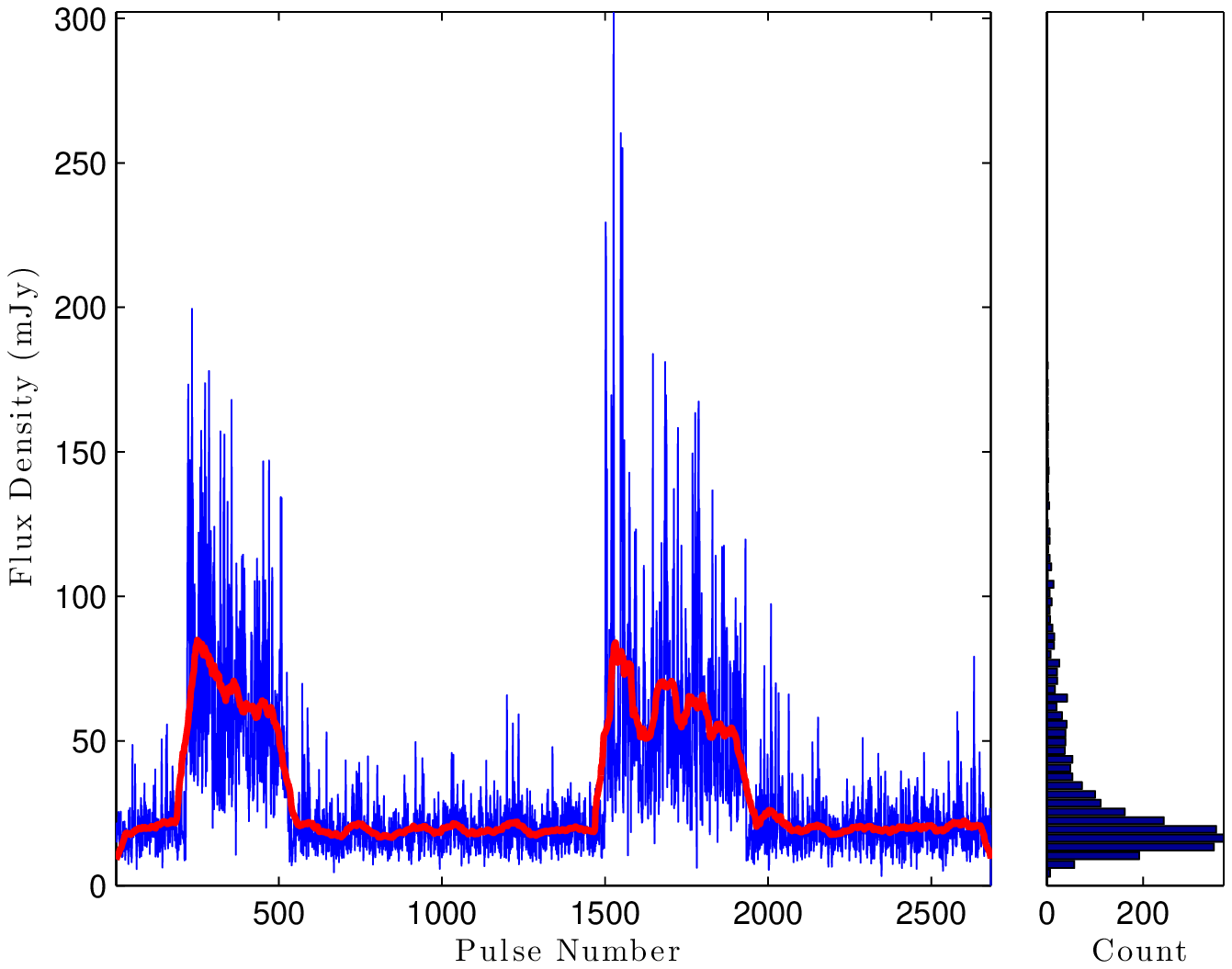}}
\caption{ Flux along the bin where the bursting events dominate in Fig. \ref{fig:98TBOX}.
The solid line is the boxcar convolution of the series.}
\label{fig:98Tburst}
\end{figure}

To examine the structure of these episodes, a bin was chosen in
a section where the abnormalities are the dominant feature. The flux
density and its convolution for that bin are shown in
Fig. \ref{fig:98Tburst}. There, a decaying behaviour can be seen across
the event before it abruptly returns to the normal emission
mode. This trend has not been recorded for this pulsar before but is strikingly similar to what was reported in
\cite{0004-637X-600-2-905} for PSR~J1752$+$2359.  There the authors describe
this type of event as bursting. We adopt this nomenclature and
will refer to this type of event as a \emph{bursting} emission mode.
		
\begin{figure*}
\includegraphics[width=.75\linewidth]{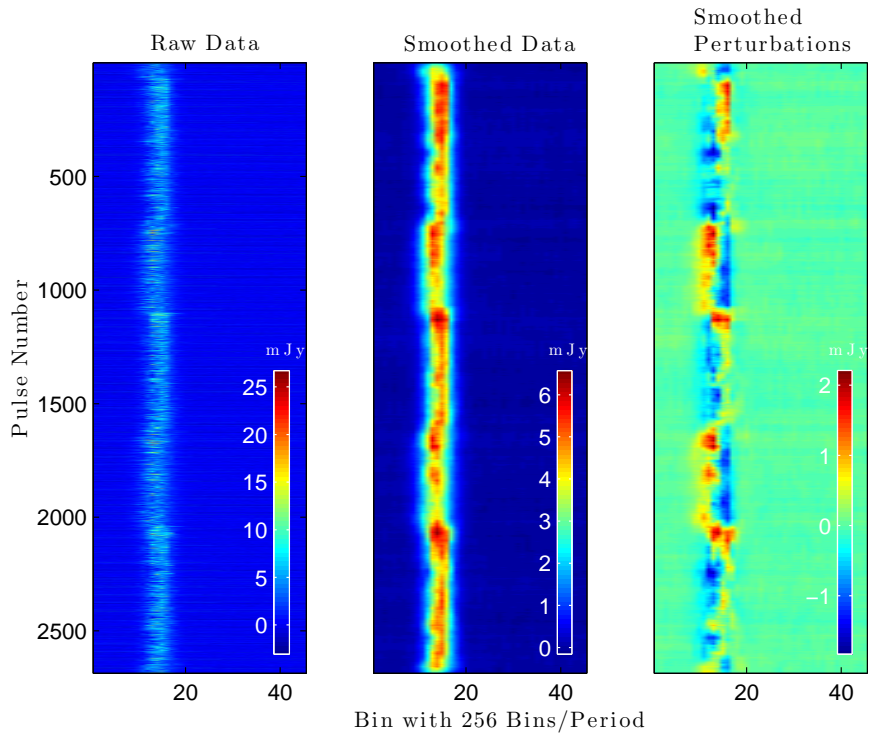}
 \caption{The 1.4~GHz observation of PSR B0611$+$22 on MJD 
 54898. Left: The folded dedispersed time series with
 one pulse profile for each pulse. Centre: 
 the time series after being box-car convolved along each bin with 
 a kernel of 64 pulses. Right: the convolution of the perturbations around the 
 mean profile with the same kernel.}
 \label{fig:98LBOX}
\end{figure*}

To investigate whether these modes can be seen at 
other frequencies, the same analysis was
conducted on data collected at 1.4~GHz. From the perturbations about 
the mean at this frequency, seen 
in Fig.~\ref{fig:98LBOX}, it becomes 
clear that there is an undulating relationship. This suggests that 
mode switching is present. 		

\begin{figure}
	\centering
	\begin{tabular}{c} 
		{\includegraphics[width=.9\linewidth,trim=25 0 100 0,clip=true]{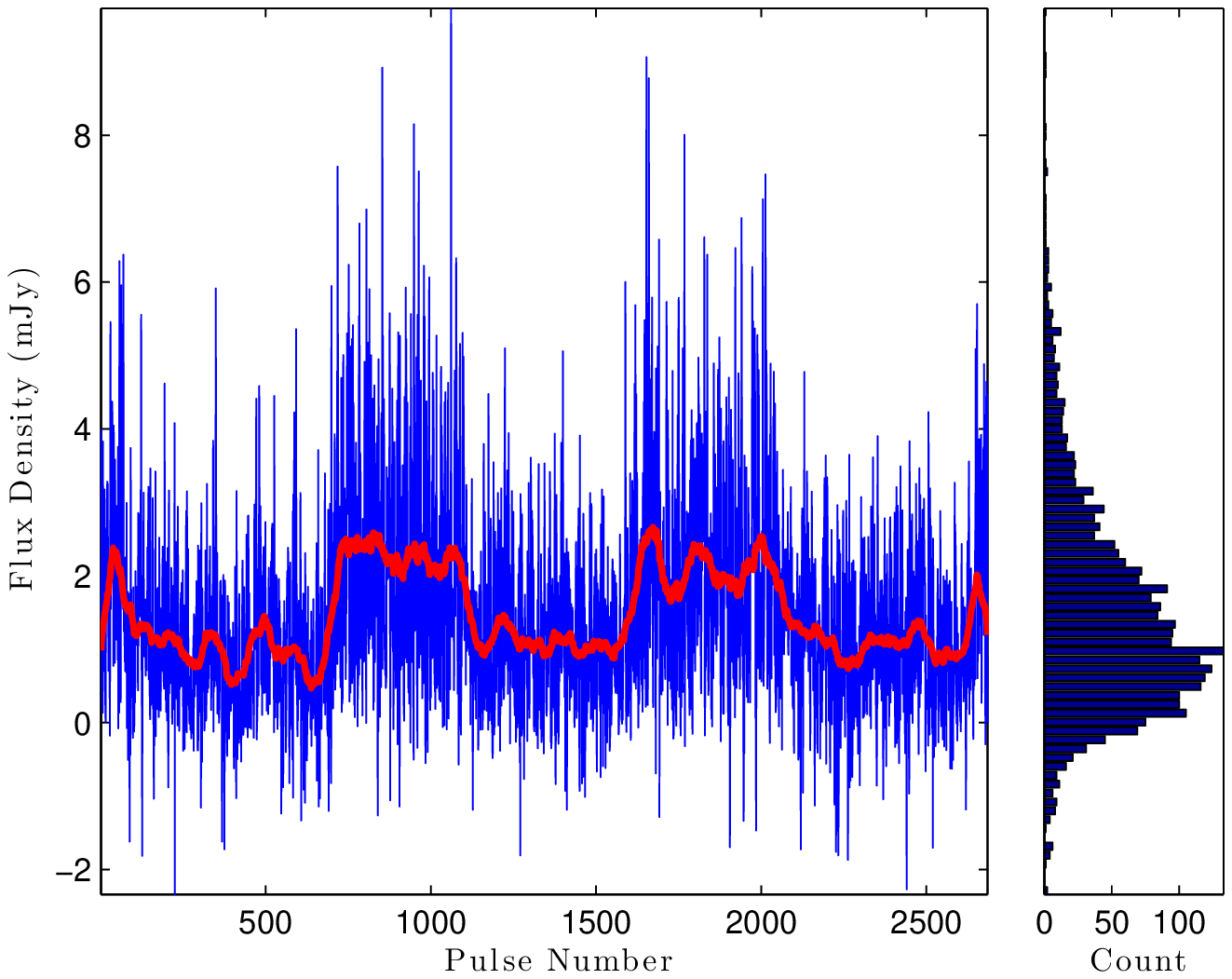}}\\
		{\includegraphics[width=.9\linewidth,trim=25 0 100 0,clip=true]{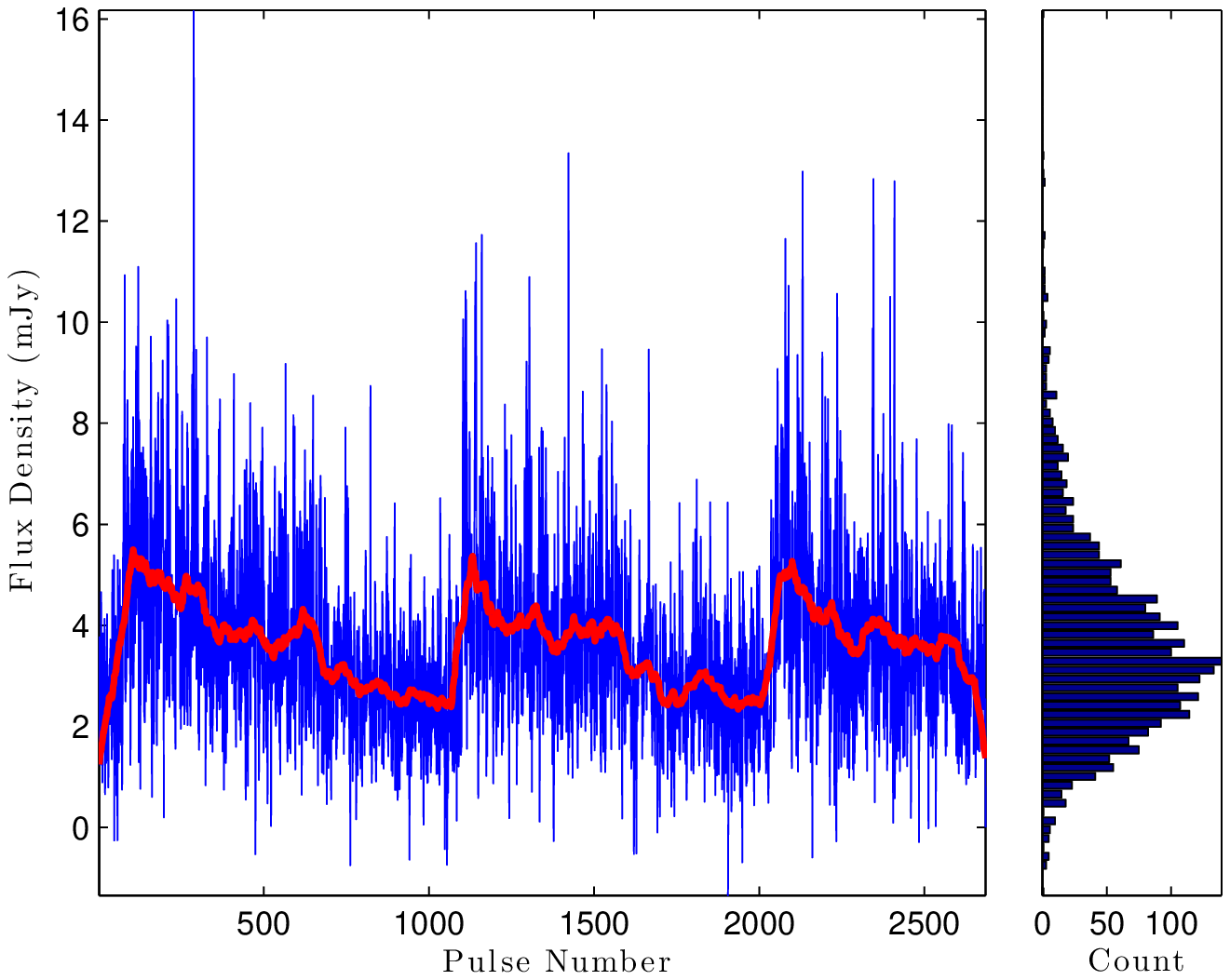}}\\
	\end{tabular}
	\caption{A 15~minute 1.4~GHz observation of PSR~B0611$+$22 on MJD 
	54898. (\emph{Top}): The flux within a bin where the left-sided 
        fluctuations dominate in Fig. \ref{fig:98LBOX}. 
        (\emph{Bottom}): The flux within a bin where 
	the right sided fluctuations dominate. 
        The solid line in both is the boxcar 
	convolution of the time series. Note the alternating flux densities between 
	the left and right sided bins}
	\label{fig:98LModes}
\end{figure}	
		
To examine whether the structure of these modes are similar to the ones at 327 MHz, we plot bin values from
either side of the undulation which are shown in Fig. \ref{fig:98LModes}. On the left side of the
undulation, shown in the top of Fig. \ref{fig:98LModes}, pulse sequences of larger flux density 
are mesa (i.e.~table-top) shaped with a quite constant mean value throughout before quickly
returning to a baseline value.  On the other hand, the right-sided
enhanced sequences, shown in the bottom of Fig. \ref{fig:98LModes}, 
appear to have the decaying structure of the bursting mode. 

This therefore confirms that the two modes found at 327 MHz are also present at 1.4 GHz, 
but with intriguing differences. Most notably, at 1.4 GHz the bursting mode in the right hand 
bin is no longer accompanied by a steady weaker mode in the left hand bin. In top of Fig.~\ref{fig:98LModes} 
we can see that the left bin is far from steady and at this higher frequency it exhibits abrupt changes of its own. 
Moreover, the left hand bursts alternate abruptly with those of the right shown in 
bottom panel of Fig.~\ref{fig:98LModes}. Together these form the undulating pattern in 
Fig.~\ref{fig:98LBOX}.

\subsection{Gaussian fitting and skewness measurements}

To quantify the bursting phenomenon, knowing that the pulse
is well described by a Gaussian \citep{1999ApJS..121..171W},
we fit Gaussians to each profile on data values greater than
25\% of the maximum value of each local mean pulse. This is done to
ensure that the fit conforms around the peak of the pulse, and to
reduce any asymmetric influences. The resulting fit parameters
(peak location $\mu$, amplitude $A$, and pulse width $\sigma$) of each pulse can then
be compared to see if changes occur between the two emission modes.  

To quantify any asymmetry in the data, we measured the \emph{skewness} of each pulse.
This procedure is outlined in  the Appendix.
In this calculation, to avoid contamination by noise, only flux density 
levels greater than 5\% of the maximum pulse value are used.

\subsubsection{Gaussian and skewness results }	
	
Note that, due to the smoothing of
the data, any structures on scales smaller than 64 pulse numbers in
these results must be treated as a noise feature. For this reason,
histograms are generated for each parameter value in order to
investigate broader statistical trends.
		
\begin{figure*}
	{\includegraphics[width=\linewidth,trim= 100 50 100 25 , clip=true]{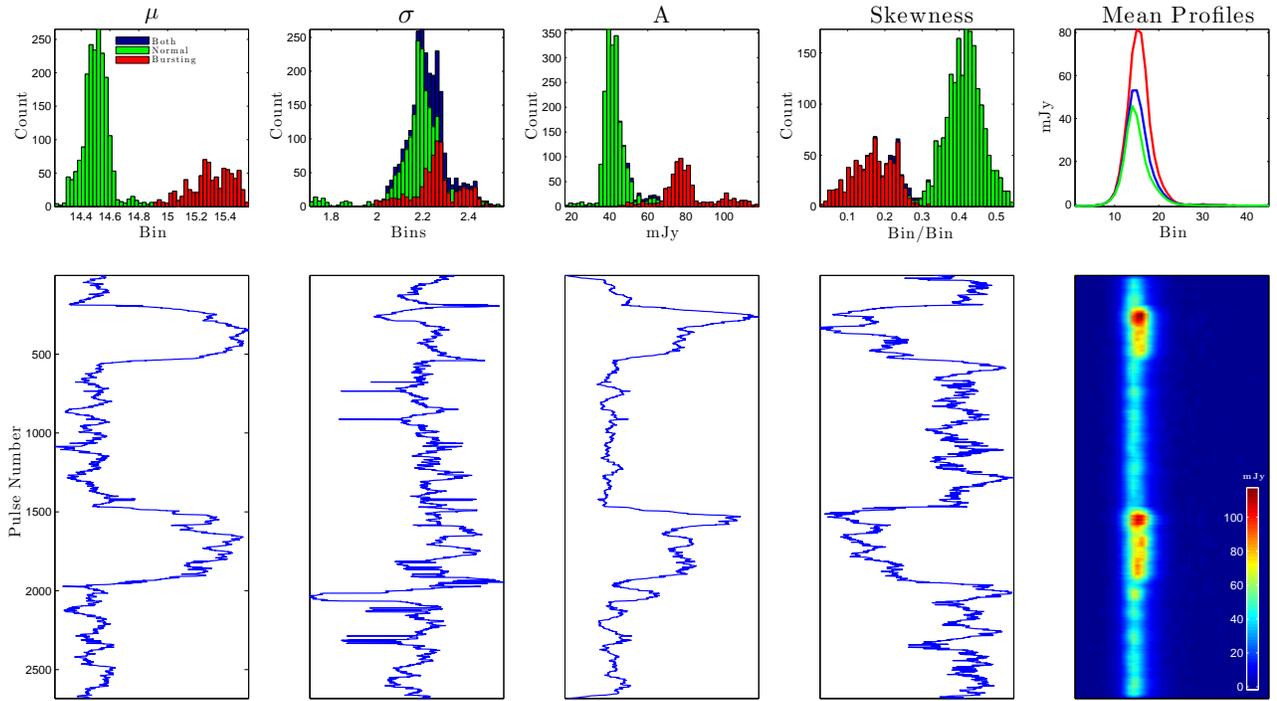}}
	\caption{Gaussian parameters and skewness measurements of the smoothed data from the 
	observation carried out on MJD 54898 at 327 MHz.}
	\label{fig:HistT}
\end{figure*}
		
The results from this analysis of the 327~MHz data are shown in
Fig. \ref{fig:HistT}. There, a clear bimodal distribution is seen in
the peak locations ($\mu$) histogram. These values are then plotted against pulse
number, the lower graph in the $\mu$ column of Fig. \ref{fig:HistT}.
It then quickly becomes evident that the larger values of $\mu$ are
associated with the bursting pulses. This directly conveys a
phase-shift between the two emission modes.
		
To isolate the two populations, we set a threshold around the midpoint (at bin 14.9)
on the $\mu$ values.  Pulse numbers with $\mu$ values below this
threshold are labeled as a normal pulse, and the rest are labeled as a
bursting pulse. From this indexing, sub-set histograms are formed for
the other recorded parameter values to see how these emission modes
contribute to each distribution.
		
For the pulse width ($\sigma$) 
values measured in the 327 MHz data, there is little that distinguishes
the two modes, while the amplitude values ($A$) are dominated at the
higher end by the bursting mode. This is congruent to what was seen in
the smoothed data set for this frequency, seen in
Fig. \ref{fig:98TBOX} and again in the far right of
Fig. \ref{fig:HistT}. When these amplitude values are plotted against
the pulse number, the lower graph in the $A$ column of
Fig. \ref{fig:HistT}, the decaying behaviour mimics what was seen in
Fig. \ref{fig:98Tburst}. This confirms that this decay is an overall 
effect on the pulse and not a single bin phenomenon.
Skewness measurements at 327 MHz show a clear separation between the
two modes, and the bursting pulses tend to have a lower skewness
than the normal mode.
		
We also wish to see if these trends are apparent in the pulse profiles. To do 
this, pulses from the raw data were averaged over the pulse numbers
of the corresponding modes and over the whole observation for
comparison. These profiles are shown in Fig.~\ref{fig:HistT}.  There
we can see that the mean bursting pulse is indeed larger in amplitude,
similar in width, and its peak is slightly shifted compared to the
mean normal pulse.
				
\begin{figure*}
	{\includegraphics[width=\linewidth,trim= 100 50 100 25 , clip=true]{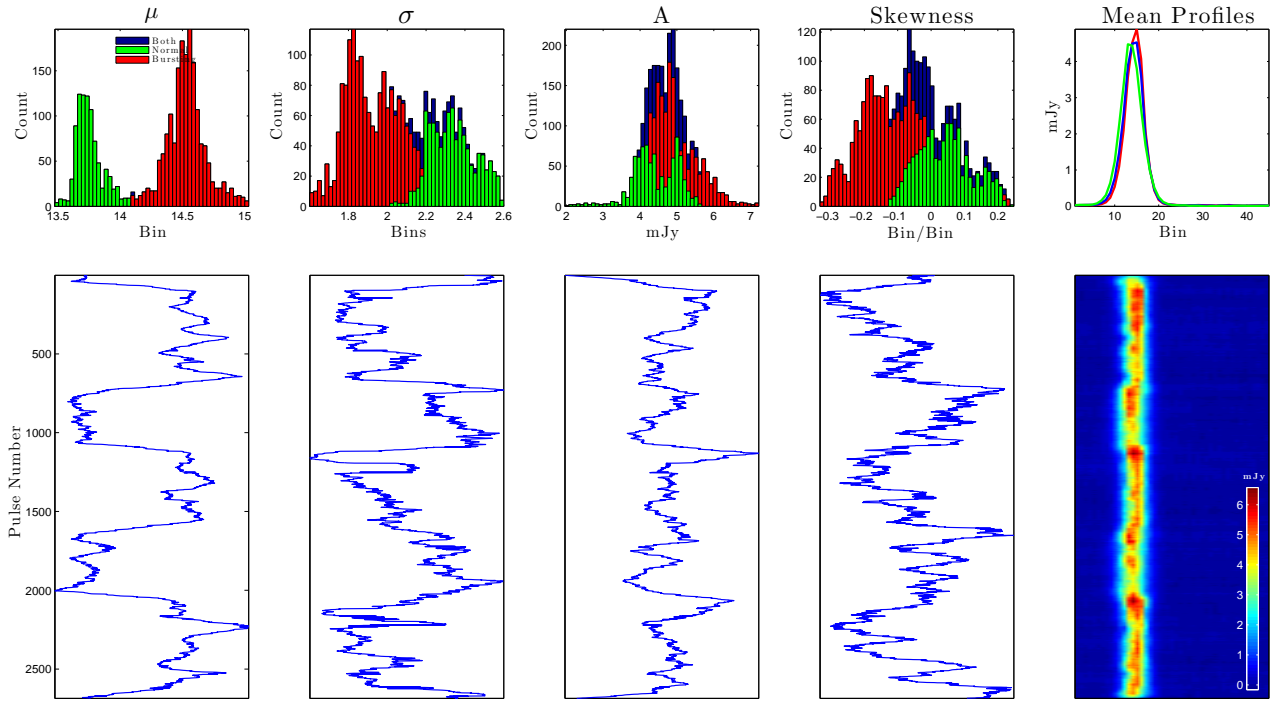}}
	\caption{Gaussian parameters and skewness measurements of the smoothed  data from the observation carried out on MJD 54898 at 1.4~GHz.}
	\label{fig:HistL}
\end{figure*}
		
To see if these trends are consistent at other frequencies, the
analysis was performed on the 1.4~GHz observation from the same day, 
shown in Fig. \ref{fig:HistL}. Again we see a clear bimodal
distribution in the $\mu$ values. When plotted against pulse number, a
noticeable phase shift is seen that corresponds with the undulation
observed in the smoothed perturbations of Fig. \ref{fig:98LBOX}. Again, a
$\mu$ threshold about the midpoint (at bin 14.1) was set to isolate the two modes for
comparison.
		
When these $\sigma$ values are compared, it is seen that the bursting
pulses tend to be narrower than the normal pulses. This trend was not seen
in the 327~MHz data, along with little distinction in the amplitude. On the
other-hand, the bursting pulses still have the tendency towards lower
skewness values.
	
	\begin{figure*}
		{\includegraphics[width=\linewidth,trim= 100 50 100 25 , clip=true]{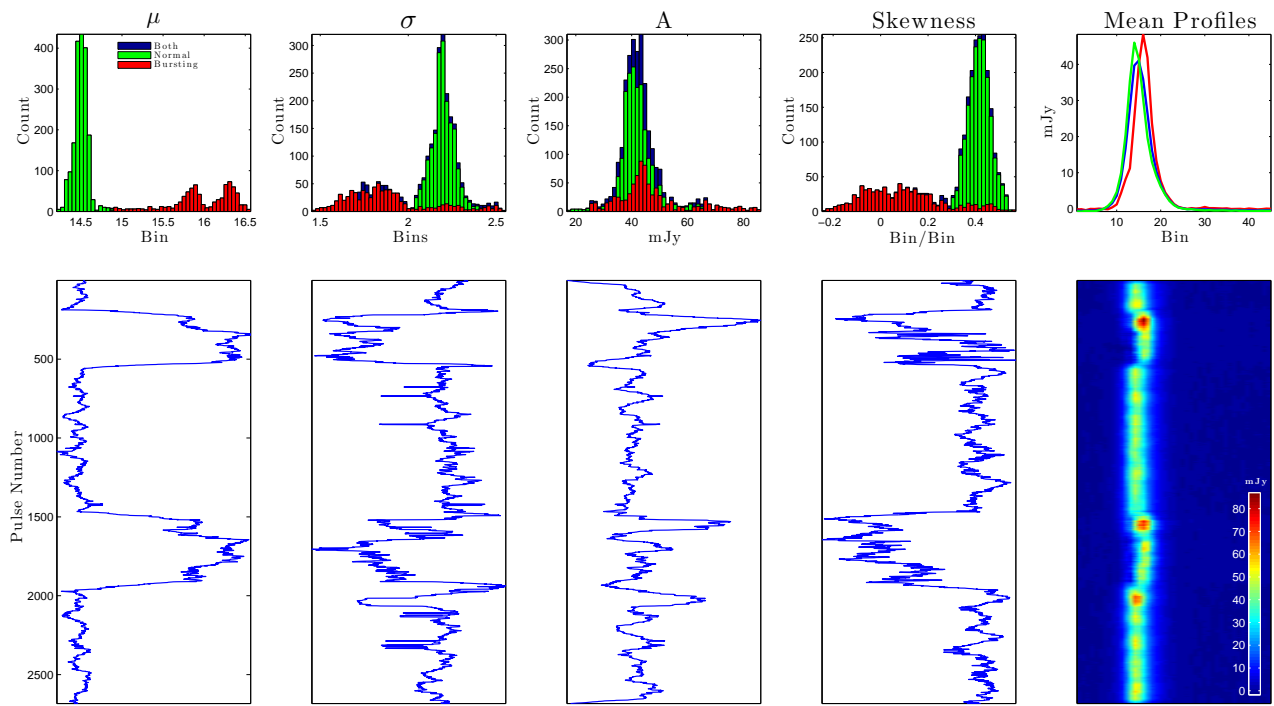}}
		\caption{Gaussian parameters and skewness measurements of the smoothed data from the observation carried out on MJD 54898 at 327 MHz with the normal profile removed from the bursting modes. }
		\label{fig:HistNR}
	\end{figure*}
		
In efforts to explain these varying trends, the 327~MHz observation
is re-examined. The larger amplitude and similar width of the bursting pulse 
allows for a nested normal pulse to be simultaneously emitting. From the indexing 
of the previous analysis, we remove the mean normal-pulse for the bursting sequences in
the raw data set. This new set is then convolved and Gaussian fitted to form 
Fig. \ref{fig:HistNR}. There we can see that the new trends reflect the ones observed at 1.4~GHz.
Where the bursting pulses are narrower than the normal mode, and the amplitudes are of 
similar magnitude.

Mean profiles for each situation are presented in the far right hand corners of
Fig. \ref{fig:HistT}, \ref{fig:HistL}, and \ref{fig:HistNR}. The bursting profiles for each are
contained within the normal profile's starting and ending
envelope. Therefore, it appears that the different emission mechanisms
are confined to a single emission cone. This confinement is also supported by 
the skewness measurements, where the asymmetry of the profile is changing to 
accommodate for this restraint, as well as the phase changes.

\subsubsection{Bursting rate}

When the analysis from the previous section is performed on the other observations from 
Table~\ref{tab:Info}, we find that bursting events are occurring rather frequently in this
pulsar, with at least one mode shift per observation. All the observations support the same
statistical trends in Fig. \ref{fig:HistT} and \ref{fig:HistL} for their respective bands. 
From the length of each burst and  
the total number of events over all the observations, we estimate a bursting event occurs approximately every seven minutes (1200 pulses) lasting for two to four minutes (300 - 600 pulses).  

\subsection{Pulse energy distributions}

\begin{figure*}
\centering
\mbox{ 
	\subfigure[]
	{\includegraphics[width=.46\linewidth,trim= 20 10 40 10 , clip=true]{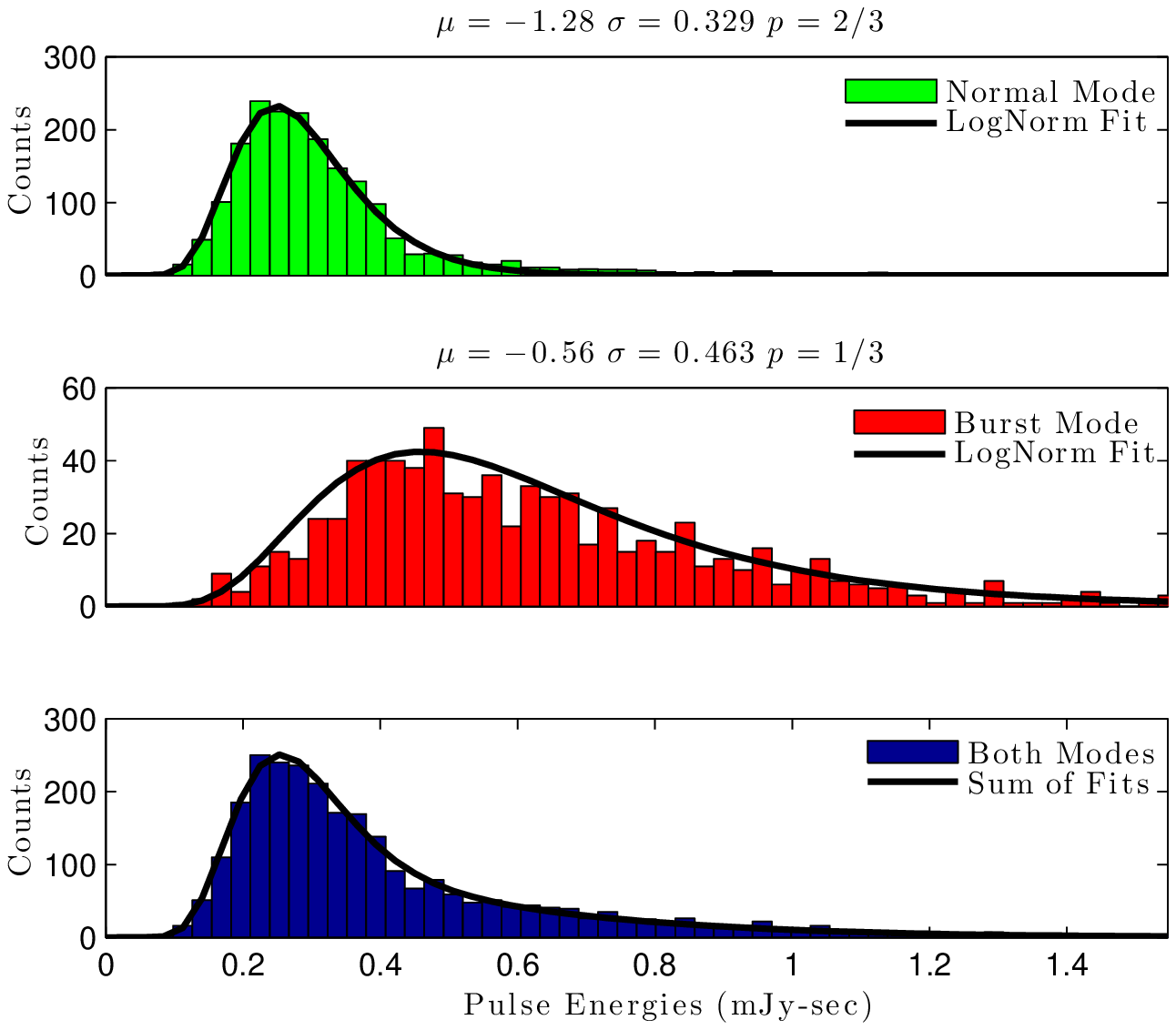}}
	\quad
	\subfigure[]
	{\includegraphics[width=.5\linewidth,trim= 20 10 40 10 , clip=true]{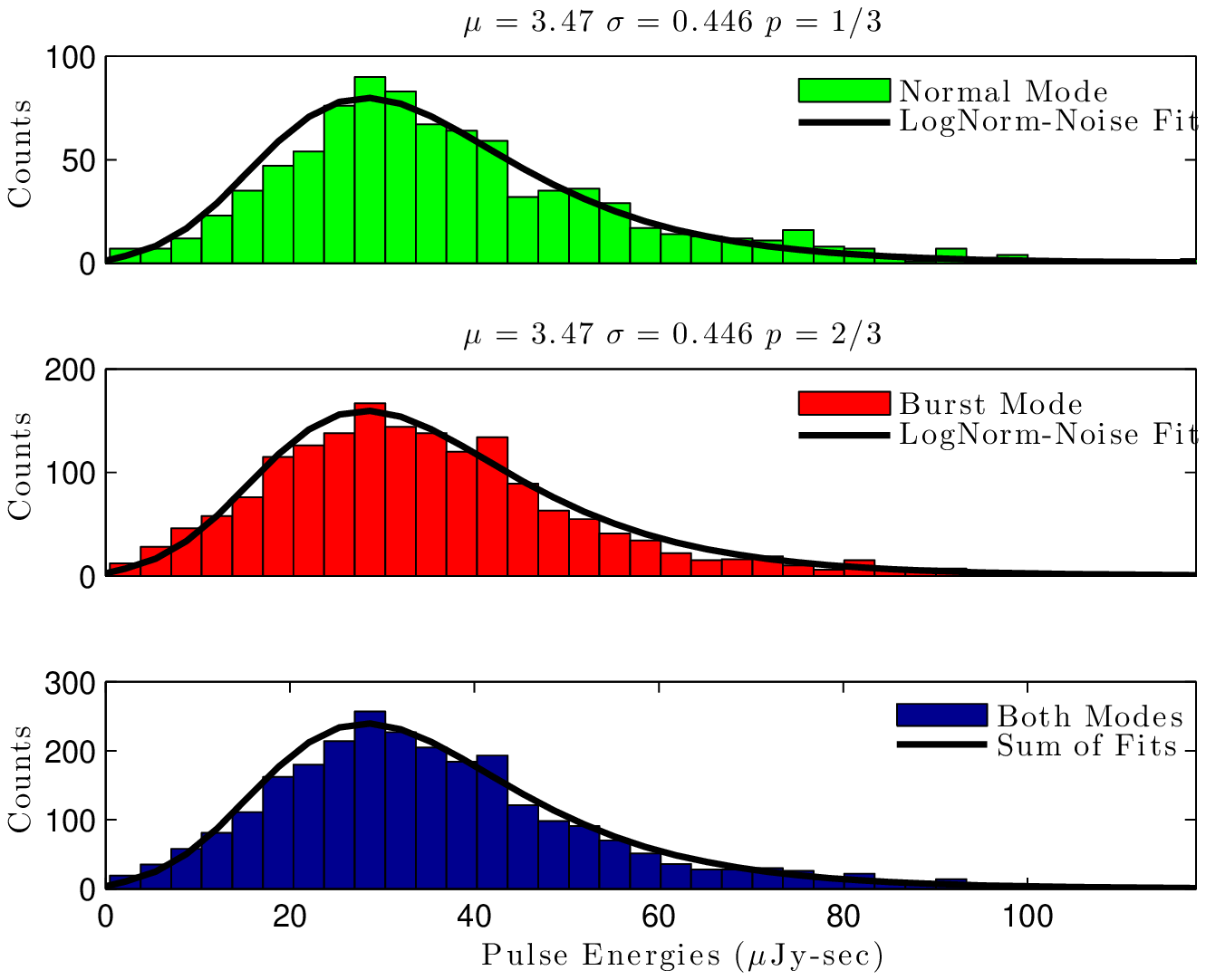}}
	}
	\caption{Pulse energy distributions for each mode and for entire observation.(a) The distributions for the
	327~MHz observation on MJD 54898 with log-normal fits. (b) The distributions for the 1400~MHz observation on MJD 54898 with log-normal effected by noise fits.}
	\label{fig:EnDis}
\end{figure*}

From the indexing in the previous sections, we are able to form sub-set pulse energy distributions for each mode.
This was done by summing ten phase bins on either side of the maximum mean phase bin for each pulse in the raw data sets. 
A bin span was then converted to time and multiplied by the sum to produce an 
energy for each pulse . The energy distributions for each mode and the overall observation are then
generated, see Fig. \ref{fig:EnDis}. 

For the 327 MHz pulse energy distributions, a log-normal probability density function (PDF) is 
fitted for each mode with a maximum-likelihood estimation (mle).
From these fits, shown in the left side of Fig.  \ref{fig:EnDis}, both modes are will described by different log-normal distributions
occupying different percentages $(p)$ of the overall observation. Here we again see that the bursting
pulses are more energetic, dominating the higher energies while the normal mode pulse are centralized
at lower energies. When these two fits are summed and compared to the overall energy distribution, 
lowest left panel in Fig.  \ref{fig:EnDis}, we can see that it matches very well to support that our indexing
separates these two mode effectively.  
 
When the same method was applied to the 1400 MHz observation, the log-normal fits did not provide a satisfactory 
fit to the data in the lower
energy range. Because of this and the overall lower energy levels, we wanted to investigate whether the noise 
was effecting an underlying log-normal distribution. To incorporate this noise to a log-normal PDF, we needed to 
integrate the probability of other values ($x^\prime$) being read at another location ($x$) due to the Gaussian noise. 
Resulting      
\begin{equation}
	PDF(x)=\int_0^\infty \frac{1}{x^\prime\sqrt{2\pi\sigma^2}}\mathrm{e}^{\frac{-(\ln{x^\prime}-\mu)^2}{2\sigma^2}}\frac{1}{\sqrt{2\pi\sigma_n^2}}\mathrm{e}^{\frac{-(x-x^\prime)^2}{2\sigma_n^2}}\,\mathrm{d}x^\prime,
	\label{eq:PDF}
\end{equation}
where $\mu$ and $\sigma$ are the shape parameters for the underlying log-normal and
$\sigma_n$ is the standard deviation of the Gaussian noise.    
Because of the symmetry of the Gaussian, this is the same as convolving the log-normal 
with a normal distribution. 

\begin{figure} 
   \centering
   \includegraphics[width=\linewidth,trim= 20 110 30 100 , clip=true]{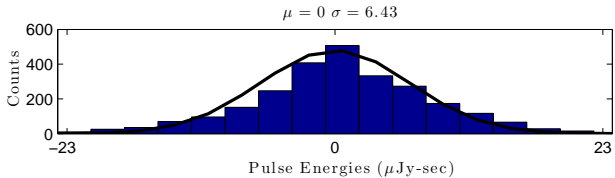}
   \caption{Off pulse energy distribution with a Gaussian fit.}
   \label{fig:Noise}
\end{figure}

To find a value for $\sigma_n$, we summed and converted 
21 bins in the off-pulse region, then fitted a Gaussian to this energy distribution, shown in Fig. \ref{fig:Noise}. 
We repeatedly found that the energies in the off-pulse regions were well described with $\sigma_n =6.43$ $\mu$Jy-sec. 

Unfortunately, the PDF in Eq. \ref{eq:PDF} has no analytical solution. Therefore, we
approximate this with an finite intervals summation and use these values as our new PDF for the mle fitting.
These fits matched the distribution remarkably well, now describing the whole distribution with little change 
in the log-normal shape parameters, shown in the right side of Fig. \ref{fig:EnDis}. 

There we can see that both modes can be described with at single energy distribution. 
This supports  what was seen in the previous sections, that there was little change in the amplitudes of 
the pulses between the two modes. What is peculiar, is that the pulse width changes between the 
modes are not significant enough to be reflected in the energies. This maybe because of the 
noise contribution.

\subsection{Sub-pulse drifting analysis}

We now take the opportunity to revisit the issue of the presence of any sub-pulse drifts in the two
emission states of B0611+22. If sub-pulse drifts are 
occurring with regular frequency, there should be a dominant peak in the Fourier transform 
of each phase bin of the raw data \citep{1970Natur.228..752B,1970Natur.227..692B}. 
While this analysis has been performed on this pulsar before in \cite{2006A&A...445..243W,2007A&A...469..607W} with 
no features found,
the two modes have not been looked at independently. Therefore, we investigate each mode separately to see if any differences
can be seen in the frequency domain.  Because we are only interested in the fluctuations of the profile, 
the mean value of each phase bin is subtracted before calculating a fluctuation spectra. 

\begin{figure*}
\centering
	
		{{\includegraphics[width=\linewidth,trim= 130 45 330 50 , clip=true]{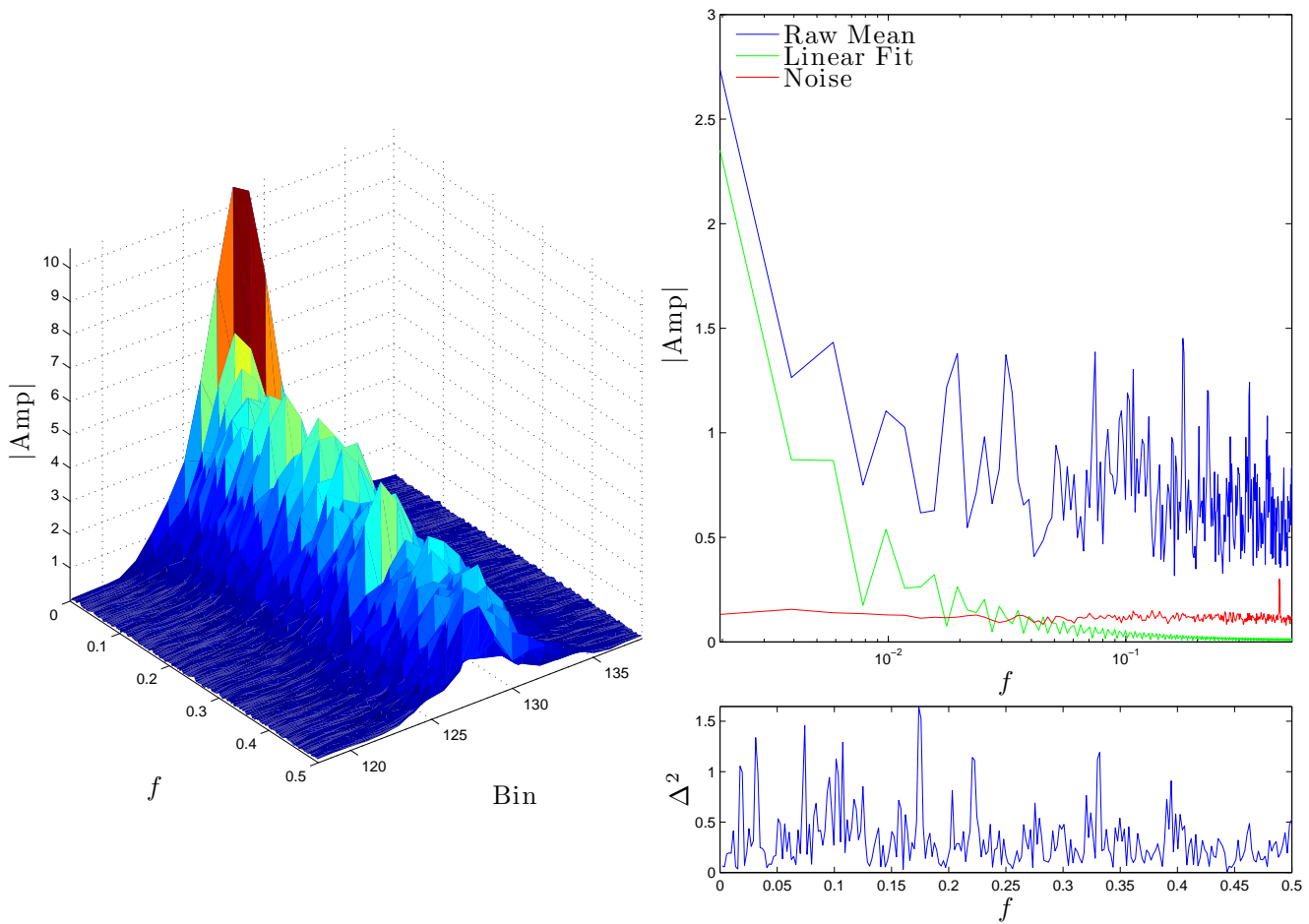}}}
		
	\caption{ (Left) The fluctuation spectra of the first bursting event for 327 MHz in Fig. \ref{fig:98TBOX}, covering 21 bins centred on the mean peak and from pulse-numbers 220 to 530. (Upper Right) The mean amplitude of each frequency of the \emph{Left} figure. Along with the Fourier transform of a linear fit of the mean flux from the same section in Fig. \ref{fig:98TBOX}, and the mean amplitude of frequencies from an off pulse area of the same size.  (Lower Right) The squared of the difference of raw mean amplitude from the amplitudes of the linear fit and the mean off source. Here all frequencies ($f$) are in cycles-per-period (C/P). }
	\label{fig:B_FFTs}			
\end{figure*}

\begin{figure*}
\centering
	
		{{\includegraphics[width=\linewidth,trim= 130 45 330 50 , clip=true]{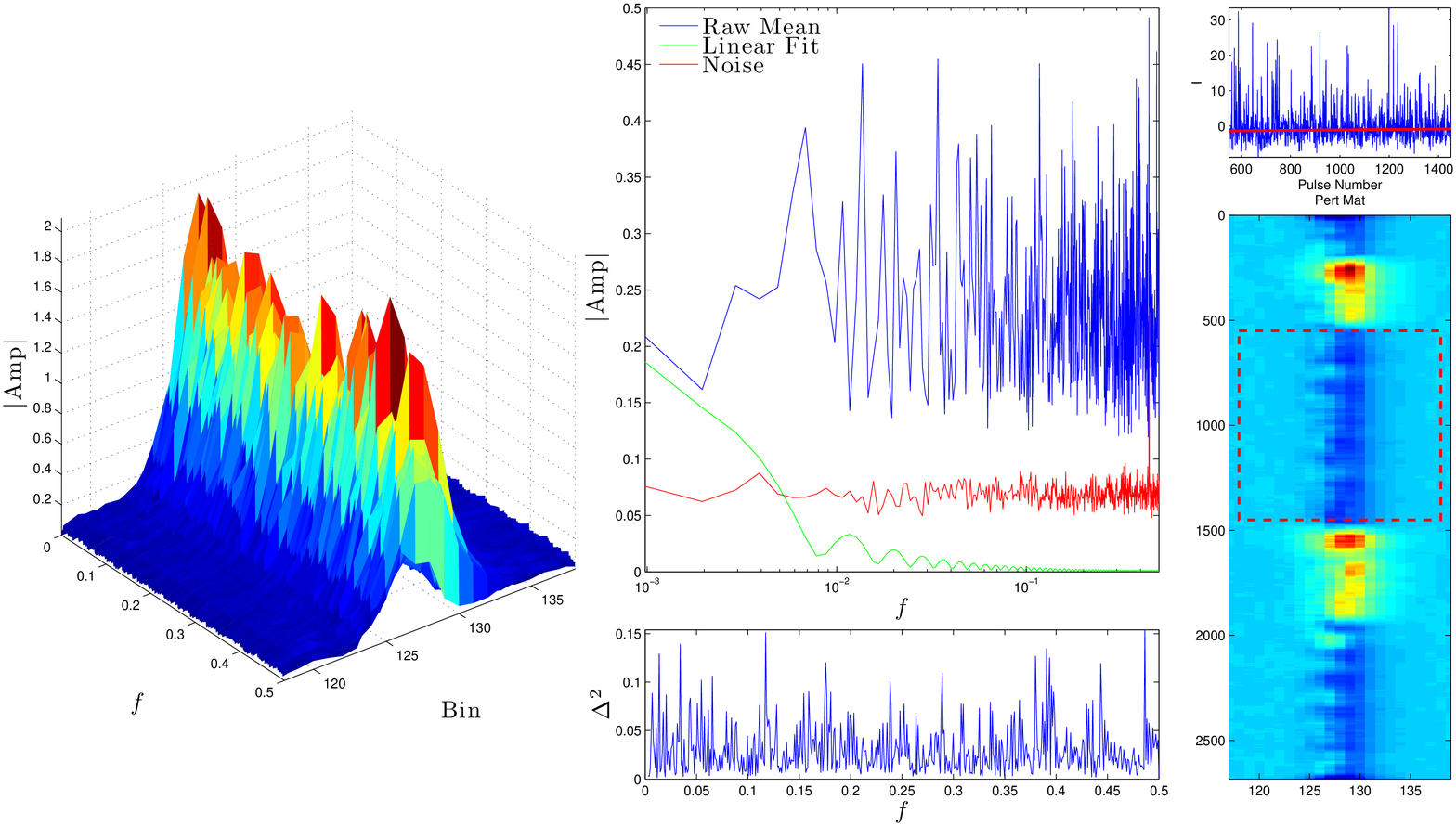}}}
		
	\caption{(Left) The fluctuation spectra of the longest normal mode for 327 MHz in Fig. \ref{fig:98TBOX}, covering 21 bins centred on the mean peak and from pulse-numbers 550 to 1450. (Upper Right) The mean amplitude of each frequency of the \emph{Left} figure. Along with the Fourier transform of a linear fit of the mean flux from the same section in Fig. \ref{fig:98TBOX}, and the mean amplitude of frequencies from an off pulse area of the same size.  (Lower Right) The squared of the difference of raw mean amplitude from the amplitudes of the linear fit and the mean off source. Here all frequencies ($f$) are in cycles-per-period (C/P).}
	\label{fig:N_FFTs}			
\end{figure*}

When investigating amplitudes of the first bursting event at 327 MHz, shown in the left hand 
side of Fig. \ref{fig:B_FFTs}, it soon becomes evident that the lower frequencies are playing 
a dominant roll. To examine why this is, we first produce a linear fit of the mean intensities of 
each pulse number. This fit is then Fourier transformed and compared to the the mean amplitude 
of each frequency in the data, shown in the upper right hand side of Fig. \ref{fig:B_FFTs}. We can then
see that in the low frequency range these two spectra are comparable. This  suggests that these
low frequencies are a consequence of the large scale structure of the bursting envelope. 

To see if any radio interference is contributing, the fluctuation spectra of an off-pulse region of the same area is 
taken for comparison. The mean value of this region is plotted in the upper right hand side of 
Fig. \ref{fig:B_FFTs}, where a single signal is seen at high frequency. This radio interference does not seem to 
be a major contributor. 
Regardless, this and the linear trend spectra are subtracted from the 
mean data spectra. This difference in amplitudes is then squared to approximate how the other 
components are contributing to the power, seen in the lower right of Fig. \ref{fig:B_FFTs}.  There
we can see that there are no substantial single frequency peaks, suggesting that sub-pulse drifting 
is not prevalent in this mode. 
When the same procedure is  conducted on the normal mode (Fig. \ref{fig:N_FFTs}), there is clearly no low frequency
dominance as what was seen in the bursting mode. This is consistent with the structural explanation 
since the normal mode should have no large-scale pattern. Again we see no prominent peaks in the amplitude
difference. 

These results are consistent with what was reported in \cite{2006A&A...445..243W,2007A&A...469..607W}.
This analysis does suggest however, that if  a `red noise' component in the fluctuation spectra it may be a sign of 
intrinsic structures in the intensity and should not always be disregarded as interstellar scintillation or 
receiver fluctuations \citep{Lorimer:2005fk}. 

\section{Predictions for observations at X-ray wavelengths}

Another very interesting possibility is that
PSR~B0611+22, and others like it, may be detectable as X-ray state
switching pulsars, as was recently demonstrated for the
classical mode-changing pulsar B0943+10 where non-thermal
unpulsed X-rays are observed during the bright radio emitting
phases, while an additional pulse thermal component is present
during the radio quiet phase \citep{2013Sci...339..436H}. 
A comparison between Fig.~1 of this paper and Fig.~1 of
Hermsen et al.~shows that the radio properties of PSRs~B0611+22
and B0943+10 are reversed: for B0611+22 we see a burst and
decay behavior, while for B0943+10 the bright mode gradually
increases in strength before declining rapidly to the radio-quiet
state. Correlated radio and X-ray observations would allow us to see 
the energy ranges over which this pulsar is switching between
emission states. In this section, we make some testable predictions
for future radio and X-ray observations of PSR~B0611+22.

If PSR~B0611+22 behaves in a similar way to PSR~B0943+10, then we
expect a higher level of X-ray emission, accompanied by spectral
changes, during its non-bursting radio phases. 
Although PSR~B0611+22 is more distant than PSR~B0943+10, its
higher spin-down energy loss rate\footnote{Following standard
practice (see e.g.~Lorimer \& Kramer 2004), we compute
the spin-down energy loss rate $\dot{E}=3.95 \times 10^{31} {\rm ergs}~{\rm s}^{-1}
\dot{P}_{-15}/P^3$ for the spin period $P$ in s and $\dot{P}_{-15}=10^{15} \dot{P}$.}
($6.2 \times 10^{34}$~ergs~s$^{-1}$ versus $1.0 \times 10^{32}$~ergs~s$^{-1}$)
provides excellent prospects for the detection of both thermal and
non-thermal X-ray emission.
To estimate the non-thermal X-ray flux expected from PSR~B0611+22,
$F_X^{\rm nt}$, we
assume an X-ray efficiency of 1\%, as inferred for PSR~B0943+10 from Hermsen
et al.~(2013), and a dispersion measure distance of 4.7~kpc and find that 
$F_X^{\rm nt} = 3.8 \times 10^{-13}$~ergs~s$^{-1}$~cm$^{-2}$.
To estimate thermal X-ray flux, $F_X^{\rm th}$, we use the
recent compilation of $kT$ versus characteristic age presented in
Fig.~1 of Keane et al.~(2013) to infer $kT=120$~eV for
PSRs B0611+22. Assuming blackbody radiation from a region around the polar caps of
an $R= 10$~km radius neutron star we can estimate
the thermal X-ray flux, $F_X^{\rm th}$. For the purposes of this
calculation, we have
adopted an emitting region with a radius of 2.7~polar cap radii.
\footnote{This choice
was motivated by the fact that the thermal emission observed in 
B0943+10 is consistent with a circular region of radius 370~m.
The classical (see, e.g.~Lorimer \& Kramer 2005)
polar cap radius for this pulsar, $R \sqrt{2 \pi R/(Pc)}$,
is 140~m.} At the nominal dispersion measure distance of 4.7~kpc,
we find $F_X^{\rm th}=3 \times 10^{-15}$~ergs~s~cm$^{-2}$.

These anticipated X-ray fluxes translate into reasonably
high count rates with existing X-ray instruments. For example,
using the {\tt pimms} simulation package, the expected count rates 
for the {\it XMM} PN instrument are approximately
0.05~s$^{-1}$ for the non-thermal emission (assuming
a power-law spectrum with a photon index of 2) and
0.001~s$^{-1}$ for the blackbody emission.
These count rates are very encouraging and translate
into detections of up to a few hundred photons per source.  The
different spectral properties of the thermal and non-thermal emission,
and the simultaneous radio monitoring, will allow the discrimination
between thermal and non-thermal emission for each
emission mode. For example, assuming a 30\% bursting duty cycle for
PSR~B0611+22 and the same X-ray properties as B0943+10, our count-rate
estimates translate to detections of $\sim 20$ thermal photons plus
$\sim 1000$ non-thermal photons during the non-bursting state, and
$\sim 500$ non-thermal photons during the bursts. The number of
non-thermal photons detected may be lower if the above 1\% efficiency
assumption is an overestimate.  In either case, however, the excellent
statistics provided by the thermal photons means that it should be possible
to readily distinguish between changes in the X-ray luminosity and/or
spectra in the normal and bursting radio states.
	
\section{Conclusion}

In summary, using a sensitive boxcar convolution method 
to preserve the phase resolution, 
we have found exceptional pulse patterns in the radio emissions from PSR~B0611+22 
in archival data taken with the Arecibo radio telescope.

 With this analysis, prolonged enhanced 
emissions were discovered at 327 MHz. When 
investigating the overall structure of this emission, it was shown that these are  
bursting events that abruptly appear and then systematically decay. 
These events appear to be superimposed upon a constant emission mode offset in pulse phase.
This was unexpected and is a completely undocumented relationship for this pulsar.   

 To gain further insight into these events, we searched for them at 1.4 GHz
using the same method, finding a different picture compared to 327~MHz. 
At 1.4~GHz, the bursting events were no longer bright episodes, and  the steady emission mode
was no longer present. At this frequency PSR~B0611+22 appears to be a mode switching pulsar, 
where one mode is the decaying burst and the other mode is constant when on.  
This mode switching is surprising in this pulsar, only being suggested 
briefly in the past
\citep{1992msem.coll..280N,2000AAS...19713004N}. PSR B0611+22's characteristic age is also surprising, 
because the majority of mode-switching pulsars are a couple of orders of magnitude older \citep{2007MNRAS.377.1383W}.  
This age discrepancy may suggest that bursting pulsars are a new classification  of
pulsars that are independent of nulling and moding.  

To investigate these relationships, we fitted Gaussians and measured the skewness 
of each convolved pulse.
These measurements allowed us to
confirm that there was a phase change between the 
two modes. This was used to isolate each of the mode statistics in each frequency range.
We then showed, that the parameter differences were reflected in each 
mode's mean profile. These mean profiles, along with the skewness measurements,  
support that the modes are confined to the same phase region and therefore the 
same emission cone. 

This analysis also shows that, apart from the amplitude 
difference from 327 to 1400 MHz, there was also a change in the distribution of the pulse-width
parameters.  To see what could be causing this distribution change, we revisited the 327~MHz
data and subtracted the constant underlying emission from the burst. When this was done, 
the normal mean profile from the 327~MHz bursting event distributions became consistent with the 
1.4 GHz distributions. 

An investigation of the pulse energy distributions also revealed differences between the two modes.
While both modes can be described by log-normal probability density functions, at 327~MHz 
the modes are independent, but at 1400~MHz both mode energies are in a single function. 

The two modes were also searched for drifting sub-pulses independently, where
no significant signals were present in any of the fluctuation spectra, supporting 
\cite{2006A&A...445..243W,2007A&A...469..607W} that sub-pulses are not prevalent in this pulsar.
We have also shown that if B0611+22 acts similarly to other known pulsars, it could exhibit changes 
in its X-ray luminosity between the two separate modes. PSR~B0611+22 is particularly attractive as
a potential X-ray source due to its relative youth. Correlated radio and X-ray observations
are being planned for 2014.

It is interesting
to note that PSR B0611+22 does exhibit period derivative variations
in a manner analogous to those published by Lyne et al.~(2010;
A.G.~Lyne, private communication based on Lovell radio telescope
observations). It is currently an open question as to whether these
changes are related to the bursting events we see here. Perhaps
the pulsar bursts only in one of the spin-down rate changes? Our
data, sampled on only a few days, are insufficient to answer this question. Further detailed
studies are required.
				
Before this analysis, the curious emission properties of B0611$+$22 
were not well known (Nowakowski 1992). Because
this is the first pulsar that we have tried with this analysis, this
may be a more common occurrence than we realise. In particular
this work required the unique sensitivity of the Arecibo telescope
to reveal this behaviour. Further studies of this kind are clearly necessary. 
 
On a more global scale, studies such as this can help draw relationships
between different types of modal and emission variations. Here it is easy to 
imagine a threshold situation that can relate nulling and bursting events.
Consider, for example, similar pulsars to B0611$+$22 that are either not as bright or 
are farther away such that the normal mode is undetected. In this case the 
normal mode would be interpreted as being in a null state. 
A similar argument was made for PSR B0656+14 in the context of a possible 
connection with rotating radio transients \cite{2006ApJ...645L.149W}.
Further examples of this phenomenon will allow us to build up a more
complete picture of the complex emission phenomenology of radio pulsars.

\section*{Acknowledgments} 

The Arecibo Observatory is operated by SRI
International under a cooperative agreement with the National Science
Foundation (AST-1100968), and in alliance with the Ana G.
M\'endez-Universidad Metropolitana, and the Universities Space
Research Association. 
This work was supported by grants from West Virginia EPSCoR, the Research Corporation
for Scientific Advancement and the National Science Foundation
PIRE award (OISE 0968296). We 
made
use of the facilities of the ATNF Pulsar Catalogue.
Computer resources used during the
later stages of this project were supported from a WV EPSCoR Challenge
Grant. DRL acknowledges support from
Oxford Astrophysics while on sabbatical leave. We thank Aris Karastergiou,
Maura McLaughlin, Andrew Lyne, Patrick Weltevrede, Ben Stappers and the 
referee, Geoff Wright, for useful comments.

\bibliography{B_ref}
\bibliographystyle{mnras}

\appendix
\section{Skewness Equations}
\label{sec:appendix}

\begin{equation}
{\cal S} \equiv
{\rm E}\left[\left({\frac{x-{\rm E}[x]}{{\rm std}[x]}}\right)^3\right].
\label{eq:Skew}
\end{equation}  
Here $x$ is a bin value, E$[x]$ is the mean value of 
$x$ and std$[x]$ is its standard deviation.

In this paper we treat each local mean pulse as a distribution function and use 
the flux densities ($y_x$)  for a weighted average. We define our weighting ($w_{x}$) as
\begin{equation}
w_{x} \equiv \frac{y_{x}}{\sum{y_{x}}}. 
\label{eq:Prob}
\end{equation}  
This then produces a weighted-mean value of
 \begin{equation}
	{\rm E}[f_{x}] = \sum{f_{x}w_{x}}.
\label{eq:mean}
\end{equation}  
Here $f_x$ is a function of $x$. 
Using this definition we write the standard deviation as 
\begin{equation}
	{\rm std}[f_{x}] = \sqrt{{\rm E}[f_{x}^2] - {\rm E}[f_{x}]^2}.
\label{eq:std}
\end{equation}  
The skewness is then calculated with an expansion of its definition, i.e.
\begin{equation}
{\cal S} =\frac{{\rm E}[x^3] -3{\rm E}[x]{\rm std}[x]^2-{\rm E}[x]^3}{{\rm std}[x]^3}.
\label{eq:BigSkew}
\end{equation}  

\label{lastpage}
\end{document}